\documentclass[10pt]{article}

\usepackage{amsmath}
\usepackage{amssymb}

\usepackage{graphicx}

\usepackage{cite}

\usepackage{color}
\usepackage{multirow}

\usepackage{setspace}
\usepackage[labelfont=bf,labelsep=period,justification=raggedright]{caption}

\makeatletter
\renewcommand{\@biblabel}[1]{\quad#1.}
\makeatother

\date{}

\begin{document}

\begin{flushleft}
{\Large \textbf{Patterns, entropy, and predictability of human
mobility and life} }
\\

Shao-Meng Qin$^{1}$, Hannu Verkasalo$^{2,3}$, Mikael
Mohtaschemi$^{1}$, Tuomo Hartonen$^{1}$, Mikko Alava$^{1,\ast}$
\\
\bf{1} Aalto University, Department of Applied Physics, Espoo,
Finland
\\
\bf{2} WR WIRELESS RESEARCH Ltd,  Espoo, Finland
\\
\bf{3} Zokem Ltd, Espoo, Finland
\\
$\ast$ E-mail: mikko.alava@aalto.fi
\end{flushleft}

\section*{Abstract}
Cellular phones are now offering an ubiquitous means for scientists
to observe life: how people act, move and respond to external
influences. They can be utilized as measurement devices of
individual persons and for groups of people of the social context
and the related interactions. The picture of human life that emerges
shows complexity, which is manifested in such data in properties of
the spatiotemporal tracks of individuals. We extract from
smartphone-based data for a set of persons important locations such
as  ``home'', ``work'' and so forth over fixed length time-slots
covering the days in the data-set (see also
\cite{whatisthelocation,significantlocation}). This set of typical
places is heavy-tailed, a power-law distribution with an exponent
close to -1.7. To analyze the regularities and stochastic features
present, the days are classified for each person into regular,
personal patterns. To this are superimposed fluctuations for each
day. This randomness is measured by ``life'' entropy, computed both
before and after finding the clustering so as to subtract the
contribution of a number of patterns. The main issue, that we then
address, is how predictable individuals are in their mobility. The
patterns and entropy are reflected in the predictability of the
mobility of the life both individually and on average. We explore
the simple approaches to guess the location from the typical
behavior, and of exploiting the transition probabilities with time
from location or activity A to B. The patterns allow an enhanced
predictability, at least up to a few hours into the future from the
current location. Such fixed habits are most clearly visible in the
working-day length.

\section*{Introduction}
The digitalized world is now getting entangled with the daily life
of almost everyone. This has various consequences from the trivial
ones - "being always connected" - to less obvious ones. But, for the
scientist this gives a chance to study human life in a quantitative
fashion. Aspects that become available include the mobility (person
X goes from A to B) statistics and the way a person reacts via a
phone to some external action
\cite{gonzales2008understanding,song2010limits,bagrow}. Quantitative
sociology then becomes boosted and enhanced by the possibilities the
phone-based data offers, and allows to study "motifs" and
regularities in life
\cite{eagle2009eigenbehaviors,topicmodels,farrahi2011}. A very
important dimension brought by the omnipresent phone to the modern
life is the social context and interactions it brings about.
\cite{eagle2009inferring,JPAPhones,vespignanisci,mitchellmining,Eagle:2006:RMS:1122739.1122745,castellano2009}.

Human life can obviously be seen to be both random and regular, and
its inherent complexity is manifested for instance in spatiotemporal
tracks of individuals - "mobility". We are in the following work
concerned with three fundamental questions: how to describe the
regularity of the patterns in the life of a person from day to day,
what kind of variations there are around that, and how predictable
are the features of mobility then after all? To this end, we exploit
a novel data-set of smartphone users (Methods) which allows us to
extract the locations the users are associated with and pick the
most important ones. The raw data is first pruned and filtered. The
goal is then to extract the significant locations for fixed length
time-slots - ``where was I from 8 to 9 in the morning?''
\cite{laasonen2009mining,nurmi2009identifying,p55-ahmad,PlaceSense,isaacmanplaces,kiukkonenLausanne}.
The data is thus distilled into a discrete format: at time-slot T
the user X is in location (or perhaps state) A forming thus triplets
(X,T,A).

These places or locations contain the most obvious ones such as
``home'', ``work'' and so forth, but it turns out that for both a
typical person and the whole set the location statistics follow a
fat-tailed distribution. The data in the form (X,T,A) can be with a
clustering-analysis turned into sets of days, that is typical
patterns ("working day" etc.): in a certain days the activity of X
at T is A, while in others it is B for instance. A day at work does
not completely follow the pattern of that particular person, so
noise is superimposed on the pattern. A physicist would say that a
person has "ground-states" of typical behaviors which are mixed with
fluctuations.

A physics or information theory based measure for the fluctuations
is the entropy, so that an individual person can be characterized by
the number of patterns he/she can be "classified with", and the
personal life entropy. The entropy is useful to compare when
computed before and after finding the patterns by clustering
\cite{shannon,kardar}. We discover that much of the raw entropy
comes from the fact that some individuals have quite a few regular
patterns of life rather than from real randomness.

The next natural question is whether one can predict the location
(or activity by indirect logic) of a person based on the current
one. The simple probabilistic guess "I know what this person does
typically at this time" works fairly well for some, but in general
does not do well. This can be improved somewhat by considering
separately all the sets of days (ground-states, again). However, the
empirical transfer matrices (related to pairs of the triplets
(X,T,A) and (X,T+1,B) from above) include more of the information in
the same patterns of life: the predictability is much higher
utilizing those and the further pattern-induced improvement is quite
mild. The capability to predict mobility fairly well in fact extends
quite a few hours into the future. This is somewhat surprising given
that the predictability is for each person dependent on the personal
entropy.

We also analyze as the most significant pattern the working days in
the data-set, clustering the days to find such patterns (Methods).
This is done from the same starting point of predictability, but
with the idea of trying to gain understanding on the origins of the
daily random deviations. In concrete terms, we find that for a
majority of the users in the set the working day has a typical,
personal length and a simple model suggests, that the daily changes
to this can be understood as a combination of a psychological trend
to stay longer or leave and the daily random reasons to
shorten/prolong the stay at work. This means, that there is a strong
degree of predictability again, and that it is influenced both by
personal (or by a particular job-dependent) details and personal
stochastic variations.

\section*{Results}

\subsection*{Days of phone users and the entropy}

\begin{figure}[h!t!b!]
\includegraphics[width=0.7\textwidth]{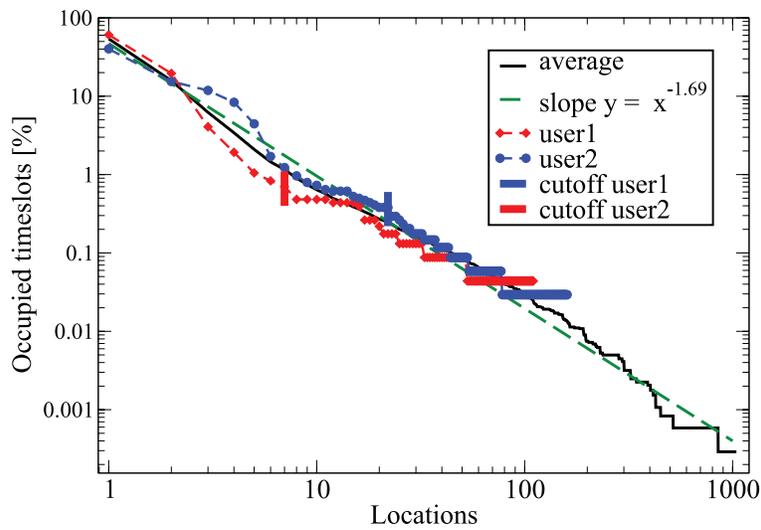}
 \caption{{\bf Important places in life}:
Average distribution (relative frequency) of locations in the
timeslots. The data is averaged over all the different users in the
set. A power-law fit to the statistics is also indicated, with an
exponent -1.69. We also show two different random users with the
``personal'' statistics plus the cut-offs (see text) for the
analysis of patterns. }
\label{fig1}
\end{figure}

The histogram of the relative time fractions that
 persons spend per location is shown
in Fig. \ref{fig1}. The distribution appears to be wide, a power law
with an exponent of about 1.7 (1.69 $\pm 0.05$); note that similar
scale-free statistics surface in studies of human mobility
\cite{brockmann2006,gonzales2008understanding} though the two
quantities are not the same. For each user in the further analysis
below a threshold is used such that locations less frequent than 1
\% of the most common one are lumped together. Two individual users
are demonstrated in the Figure \ref{fig1} including the respective
thresholds (7 and 22 significant locations). Note for instance how
the statistics for user 2 are close to the averaged behavior. It is
interesting that such regularities are found for individuals. The
fraction of locations where some time is spent, that is how many
locations are relevant, depends on the time during the day or the
slot T (Supplementary Information) \cite{lowentropynokia}.

\begin{figure*}[h!t!b!]

\includegraphics[width=0.9\textwidth]{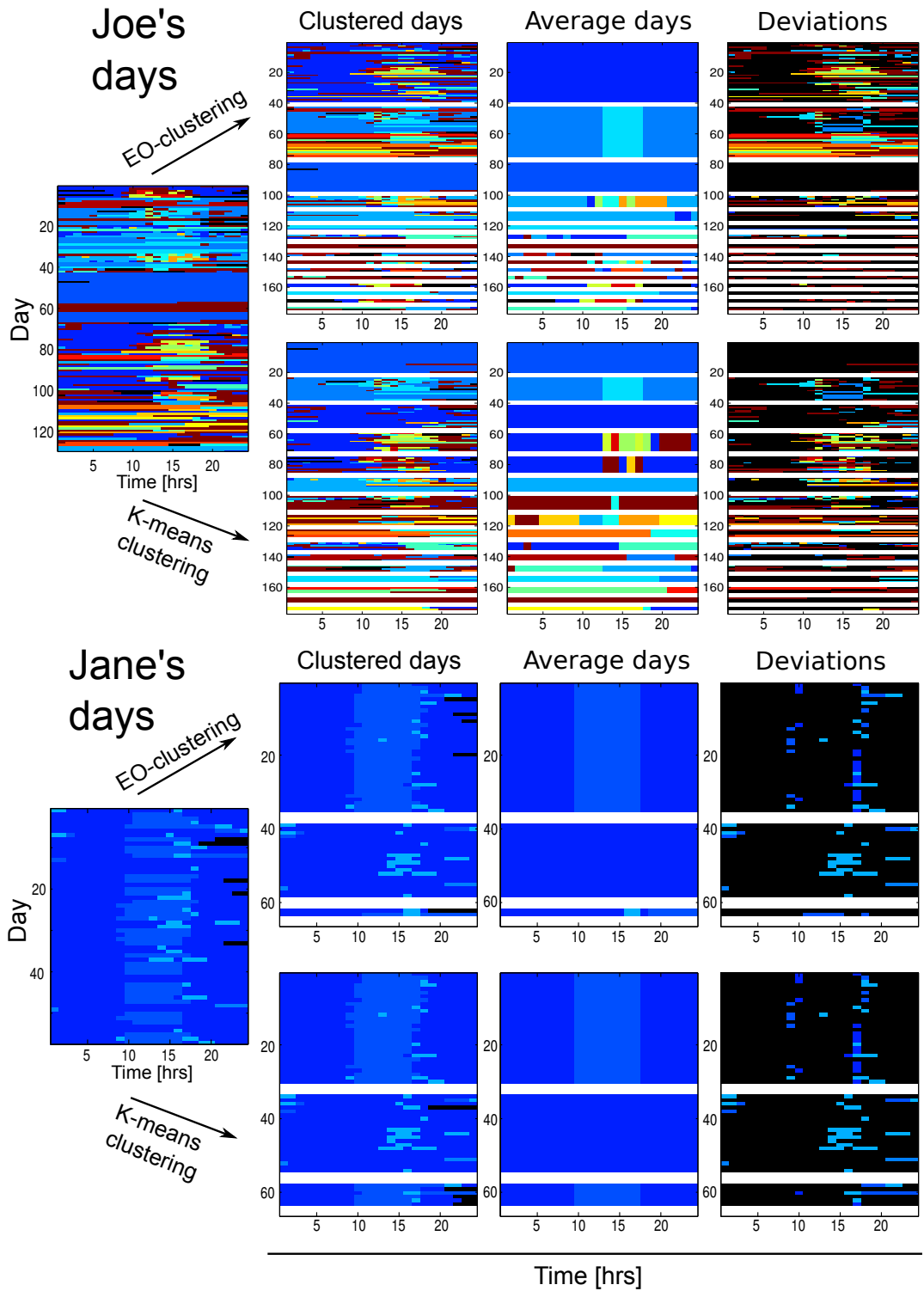} 
 \caption{{\bf Patterns of life: Joe Random and Jane Regular}: Upper panel: Clustering result for the \textit{dayvector}s of \textit{user Joe}.
 Lower panel: Clustering result for the \textit{dayvector}s of \textit{user Jane}.
 In both panels, in the first row are the results obtained with the
extremal optimization algorithm and in the second row the
corresponding result with the $k$-means clustering. Each color is
one location | the clusters are separated by white lines. The first
column shows the unclustered days. The second and third column show
the clustered days and the corresponding average days of each
cluster. The last column shows the deviations from the average day,
where the color of the location is only shown if it deviates from
the location of the average day of the cluster | otherwise the color
is black.} \label{fig2}
\end{figure*}

Two example users, say Mr. Joe Random and Ms. Jane Regular, are
depicted in Figure \ref{fig2}. The dominant patterns can be clearly
identified as candidates for a typical ``working day`` and a typical
''home day``. Two main features are the randomness or deviations
present - consider  a typical work day, where the noise is at both
ends of the period at the work location - and the similarity of the
clustering results. These two choices from extreme ends of the
spectrum of the set of users illustrate the variety of behaviors. We
also compare the results from two different clustering methods (EO
and k-means, see Methods). The simple question here is how much the
results from the analysis depend on the method applied.

 Next, we compute a personal entropy over the time slots by
\begin{equation}\label{eq:entropy_uncl}
\varepsilon= - \frac{ \sum_{j=1}^N \sum_{i \in I} l_{ij}/D_{tot}
~log (l_{ij} / D_{tot}) }{ N}
\end{equation}
where $N$ is the number of time slots in one day (columns),
$D_{tot}$ is the total number of days (rows), $I$ is the set of all
visited locations and $l_{ij}$ is the number of times location $i$
dominates time slot $j$. A low entropy means the user is regular and
vice versa. However, part of the randomness is only apparent due to
the patterns. The extreme limit is when the $l_{ij}$ split perfectly
among the different patterns (e.g. one is only at work in a certain
slot in one, and elsewhere in the others). Thus one should consider
a clustered entropy that gives in that limit the correct result,
zero. Such one is the weighted average of the single cluster
entropies:
\begin{equation}\label{eq:entropy_ave_cl}
 \bar{\varepsilon}_c=-\frac{\sum_{c \in C} \sum_{j=1}^N \sum_{i \in I}
l_{ijc}~log(p_{ijc})}{D_{tot} N}
\end{equation}
where $D_c$ is the number of days that belong to cluster $c$ and
$l_{ijc}$ is the number of times location $i$ is the most dominant
location at time slot $j$ in cluster $c$, $p_{ijc}= l_{ijc} ~ /D_c$
and $C$ is the set of clusters. The drop in entropy due to the
clustering is then defined as
\begin{equation}\label{eq:entropy_drop_percent}
\Delta \varepsilon_{[\%]}=100 ~
(1-\frac{\bar{\varepsilon}_c}{\varepsilon}).
\end{equation}
This gives also a measure for the ``goodness'' of clusterings (the
higher $\Delta \varepsilon{[\%]}$ the better). When considering
entropies one should recall that the clustering has been done a
priori, without using entropy change to guide it. One possibility
would be to look for the patterns that minimize the entropy, in
other words. For the $k$-means case, the example which minimizes the
distances to the cluster centroids is picked, which is in practice
close to taking the smallest entropy value over the trials.

\begin{figure}[h!t!b]
\begin{center}
\includegraphics[width=0.9\textwidth]{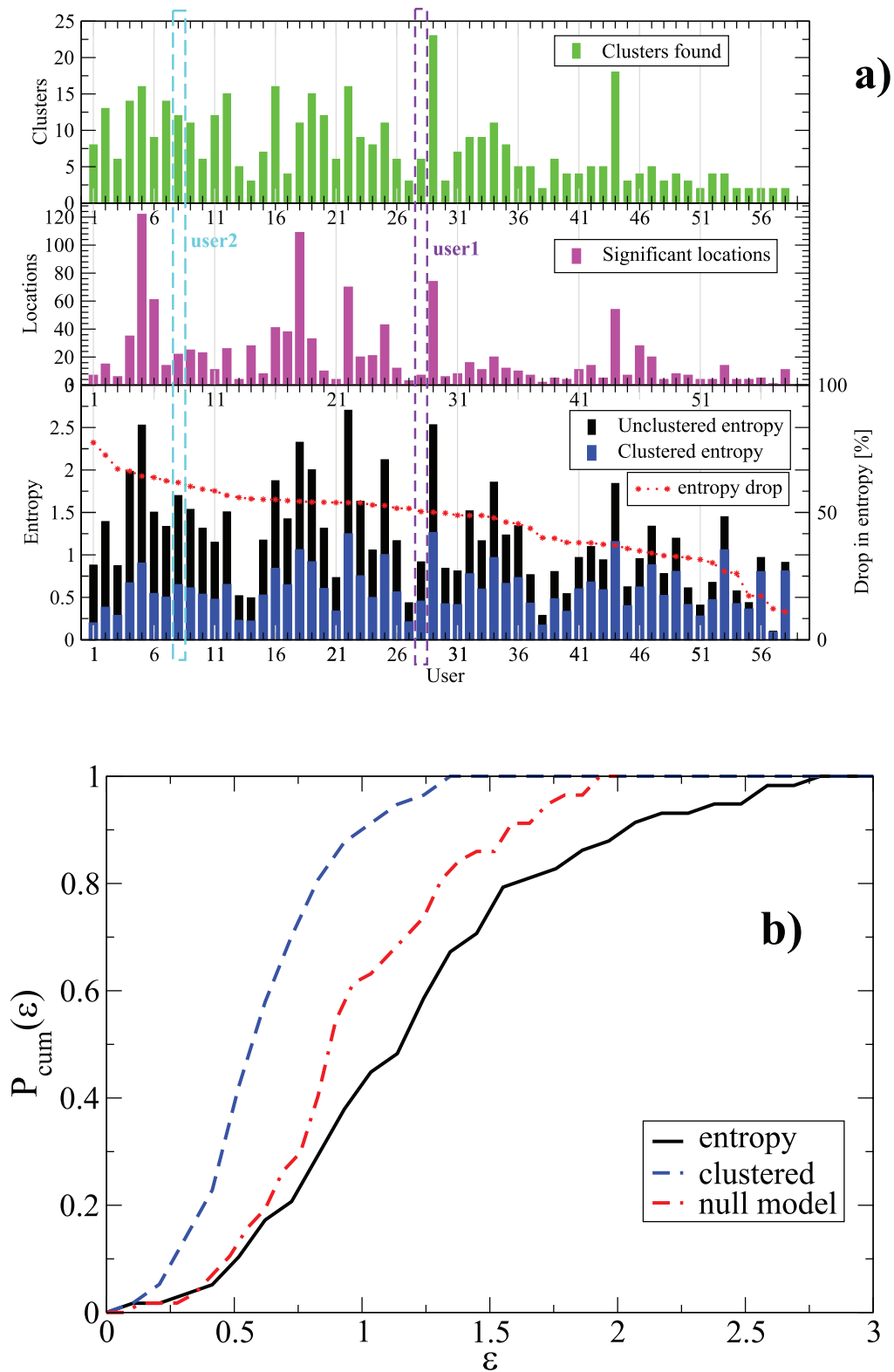}\\
\caption{{\bf Statistics of patterns, and life entropy}: a) The
clustering results for 58 accepted users with the first place
definition (off-line clustering) and a time slot of 1 hour. The
clustering was done with the EO algorithm and $\beta=10/24$. b) The
cumulative distributions of the entropies for the three cases - bare
entropy, reference model, and after clustering.} \label{fig3}
\end{center}
\end{figure}

Joe and Jane exhibit different entropies and relevant patterns. Joe
has a bare entropy of 2.70 decreasing to 1.25 and 0.78 after EO and
$k$-means clusterings, respectively (54 and 71 \% changes). Joe's
life has 16 typical patterns. For Jane (3 patterns) the entropy is
0.44, and after clustering 0.21 (53 \% decrease). The results over
the user set are summarized in Fig. \ref{fig3}a,  ordered according
to the entropy drop. A picture emerges of large entropy variations
from user to user, and a trend in the entropy with the numbers of
significant locations and patterns. The latter two fluctuate from
person to person, but not with any kind of power-law statistics as
that seen in Fig.~\ref{fig1}. There are thus two contradicting
observations: the entropies vary quite some, while the location
statistics present regularities. 
The cumulative entropy histograms are
presented in Figure \ref{fig3}b. It appears that they are quite
close to {\em log-normal} distributions, though the tail properties
can not be resolved convincingly with the histograms at hand. The
question of where do the ``rare'' locations that exist in the
location data end up or what they signify has an easy answer: a
direct comparison of the noise superimposed on the patterns reveals
that as one could expect the more rare locations are found (more
frequently) in the noise (ie. when the actual location (X,T,A) does
not agree with the pattern-indicated location (X,T,B), Supplementary
Figure 26).

We used a null-model, included in the Figure \ref{fig3}b, to compare
the entropies: assume the pattern(s) can be measured by a
concentrated probability $p_c$ for a pattern (for the descriptive
location), plus the residual variations around it in $N-1$
locations, where the $N$ is found for each user separately as above.
This means, that the $1-p_c$ part is assumed to correspond to a
completely random case. This would mean that the patterns
(descriptive location) are just mixed with noise. Then, the maximal
entropy can be written as $\bar{\varepsilon}_c = \varepsilon(p_c) +
(1-p_c) \ln(N-1)$.  We find that the distributions are ranked so
that the one for the patterned cases has the smallest entropies.
This means most likely that additional information is hidden in the
noise. The entropies for the user set do not extend in this case to
very high values. The median entropies are 1.16 and 0.56,
respectively. Such values have the simple interpretation that there
are around three relevant locations for any slot, which decreases to
less than two for a typical pattern.

\subsection*{Predictability of locations}
To which degree one can predict a single user's behavior from the
current state, ie. location A at time T? If the entropy with
patterns would be zero, this could be easy: just know the "actual
pattern" to get a perfect accuracy. Such regularities or high
predictability are analogous to what is indicated in some cases by
common sense ("at sleep during night", Supplementary Figure S27). We
consider next predictions similar to weather forecasting: first,
using the ``average'' behavior as in what is the expectation for a
slot, and second, by asking what is the probability $p_B(t+\Delta
t)$ of the location $B$ at $t+\Delta t$ if the user was at $A$ at
$t$? The second method amounts to using the measured transition
matrices and checking what they imply for the future from the
current location, comparable to the recent Markov chain idea \cite{markovpred}. Since in both cases we can also restrict the days
to one pattern at a time and average then over the patterns, we have
in effect four different methods. The recent activities on "next
location prediction" and similar topics quote a rather large number
of related methods; our results below are better than any of those
known to us from the literature
\cite{highgranular,nextlocpredperiodicity,machinelearningnextstep,nextlocpredhandwave,nextlocpredhphd05,nextlocpredmatrices,nextlocpredbayes060} though
Gambs et al. \cite{markovpred} reach comparable levels using GPS-based data.
Etter et al. \cite{nextlocpredbayes060} for instance note also the
possible roles of  data sparsity, noise, and personal variations. Of
these we analyze the two last ones below in more detail.

\begin{figure}[h!t!b!]
\begin{center}
\begin{tabular}{c@{ }}
     \includegraphics[width=0.4\textwidth]{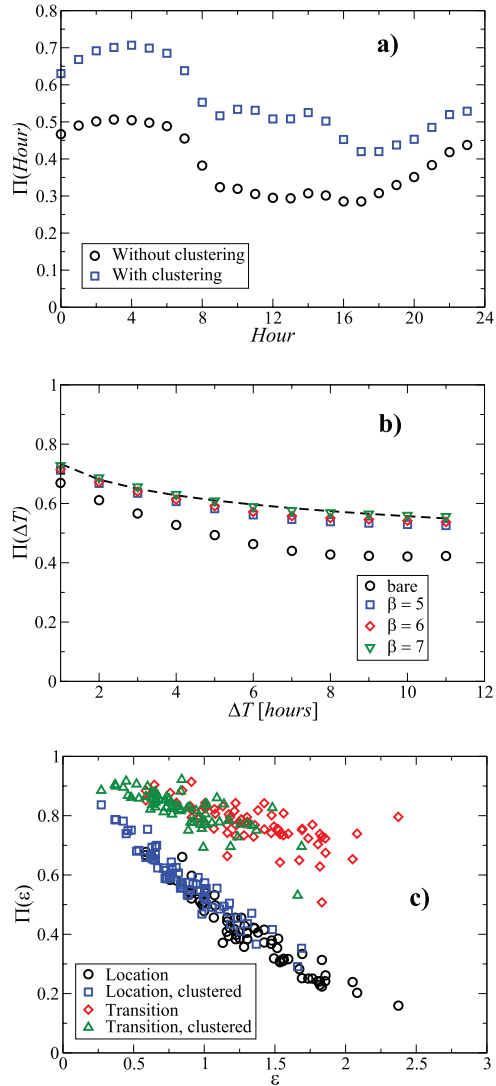}\\
 \end{tabular}

    \caption{{\bf Predicting life}: a) The prediction accuracy of the first, location based method in the
     case of with clustering and without as a function of time during the day. The clustering is from the EO algorithm with $\beta= 7$. b) The predictions by the
     transfer matrix method for various $\beta$ and $\Delta T$ up to 12 hours. The dashed line is a logarithmic fit to the $\beta =7$ case.
c) The correlation of the personal entropy and the resulting
predictability for both the prediction methods ($\beta=7$). The
entropy values are either the ``bare'' ones or those after
clustering if that is used to aid prediction.}
    \label{fig4}
\end{center}
\end{figure}

The first prediction quality is $ 
\Pi_{A(t)}$, 
user will be at time $t$. $\Pi_t$ averaged over the whole day is the
total prediction quality. With $p_{ijc}$  the probability to be at
$i$ at time slot $j$ in pattern $c$, then, quite simply
$\Pi_t=\overline {p_{ijc}^2}$, with the average over all $i$, $c$.
This method amounts to utilizing the measured probabilities to get a
a guess where a user would be - consider a two-state model where
these locations would have probabilities $p_1$ and $p_2 = 1- p_1$,
in which case such a guess works with the quality $p_1^2+p_2^2$. The
results are shown in Fig.~\ref{fig4}a. The prediction accuracy
improves with patterns to 0.53 $\pm$ 0.12 from 0.36 $\pm$ 0.11
(without).

With the second method, considering the equivalent case by setting
 $\Delta t=1$ the accuracy is much
higher. The average value changes adding the patterns from 0.74
$\pm$ 0.07 to 0.78 $\pm$ 0.06 (Fig.~\ref{fig4}b), the best result of
the four different methods tried. The small difference or change
from the patterns implies that they are already quite well encoded
in the original transition probabilities. This is natural since here
we consider the transitions (or mobility), not the static
probabilities. Thus in the case of the previous example one could
find that the two states always change to the other at the next
timestep: this has a quality of one. The maximum mean predictability
is suggested by the method and data to be close to 0.8 even with
further refinement in finding patterns, due to this small
improvement. The upper limit of the last case, with patterns, as
observed from the distribution of personal predictability values
found here, is quite close to the maximum limit postulated for the
predictability of mobility patterns, 0.93 \cite{song2010limits}.
This might not be surprising, that the predicability of daily
routine is constrained by the randomness-induced limitations in
mobility prediction.

The patterns imply that the transitions contain useful information,
which then brings us to look at long-range predictability, $\Delta
t>1$. The quality decays logarithmically with $\Delta t$,
Fig.~\ref{fig4}b, or possibly as a power law with a small exponent
(Supplementary Figure S28). We have not considered longer-range
predictions than those depicted here mostly due to data limitations.
Surely this question is interesting and has non-trivial features -
the sleep-wake cycle, the weekly cycle of work and leisure among
others. The slow decay observed or measured implies that life in its
regularity is - again given that there are such patterns - actually
relatively quite predictable. In particular, note that the quality
remains better than the average of the location method for $\Delta
t=1$, even up to time-delays of several hours. Below, we discuss the
regularities of working days, where a high level of predictability
is often quite given a priori.

Predictability has large personal variations (Fig.~\ref{fig4}c) due
to entropy if one tries predicting without exploiting the patterns.
The range in which the first type prediction quality varies shows a
substantial variation from ``easy to predict'' to ``difficult to
predict'', which correlates as stated with the entropy
(Supplementary Figures S29-S30). Joe's life would be appear to be
quite impossible to capture without patterns, but including those
increases the predictability substantially ``upto'' the level in the
case of Jane. In other words, the implication would be that we are
slaves of our daily-life habits, but those can be quite plentiful
and different and still exhibit typical deviations.

\subsection*{Departing from work}
\begin{figure}[h!t!b!]
\includegraphics[width=0.5\textwidth]{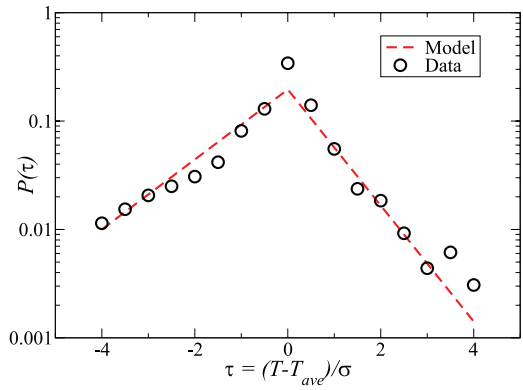}
 \caption{{\bf Predicting leaving work}: Length-of-working-day distribution in scaled units and the model
fit. The parameters for the leaving early and late -processes are
0.045 and 0.046, respectively (in inverse $\sigma$-slot lengths).}
\label{fig5}
\end{figure}

The regularity in the data includes and implies long-range daily
correlations: the behavior depends on the past. The easiest example
of this is the working day length, which we consider next. Figure
\ref{fig5} depicts the distribution for the probability to leave
from work (to any other location), for users that have regular
working patterns and enough data, slightly more than a half of the
total (Supplementary Figure S31). The data has been first scaled by
using the mean and the standard deviation for all such users so that
one is able to consider the sum distribution.

The tails of the likelihood are different on both sides of the
maximum. The Figure \ref{fig5} includes a fit of a model (see also
Supplementary Figure S32). First we assume that the time spent at
work is shorter in a particular day
 when one leaves for a specific reason
(visiting a doctor, meeting friends/family...). The action of
leaving from work has a likelihood per time step, which increases by
a constant factor by each step until the day has reached its typical
length. This assumption means that it becomes more easy to quit
working if the resulting deviation from the usual is shorter.
Sometimes on the other hand work requires extra attention. In the
model, we assume simply that if the day lasts beyond the typical
length the probability per time step to leave increases with another
parameter. This means that it becomes more difficult to stay for
each further step, a trend which is exactly the same as assumed for
the other case of leaving early.

These assumptions boil down to two exponentially increasing trends
to leave early or not to stay any longer, which can be used to fit a
distribution to the data. For the data at hand, the model works
fairly. The central assumption behind the model is that the average
working day length has for each user a deterministic influence on
the decisions to change once from the usual habit, perhaps a
triviality in many professions. This conditions the patterns over a
large time span, as seen earlier in the context of long-range
mobility prediction (compare with. Fig. \ref{fig4}b). We also tried
to correlate the departure from work with the personal communication
records. This amounts to looking at call and text message patterns
and partners (numbers for in- and out-coming communications
considered also in the light of their frequency). The idea would be
roughly: "I go since I made a call or got a call", possibly also
conditioned on the quality of the communication: either a rare or
frequent contact as deduced from the randomized phone number.
However, such attempts were left inconclusive. One reason is that
the persons considered in this study on average communicate (calls,
text messages) fairly rarely, so any possible causality can not be
derived from data. The lack of correlation in any case implies a
relatively minor role of such communications in enforcing a
particular departure from routine.
\section*{Conclusions}
To summarize, smartphone -based studies allow for quantitative
insights into the fundamentals of human life. We have here analyzed
such data with the approach of first extracting patterns and a
measure of randomness, and then analyzing the predictability of
individual mobility. This utilizes the coverage in space and time
available from the data-set. All in all, this means that such
spatiotemporal patterns, together with the personal randomness,
define quantitatively individuals and populations. Patterns and
entropy relate to the degree the daily locations or activities are
predictable. We have demonstrated a fairly good quality (0.78 in the
best of the four methods tried).

The data-set we have at our disposal is just sufficient for the
present purposes. The user panel, with its particular socio-economic
and geographic background (for a comparison, consider
Ref.~\cite{orange}) and its size may introduce various bias, which
however pales in importance if put into contrast with a few bigger
issues in the context of human life analysis. We already find
noticeable variations in the simplest personal quantitative
characteristics: the number of locations to describe the personal
life, the number of patterns, and the entropy, in the case at hand.
It appears that the consequent realistic predictability arises from
the presence of both patterns and noise in the life. Quite good
predictability - the best such to our understanding so far - is
found from the combination of using mobility and the discovered life
patterns.

One follow-up question is: how do entropy and predictability change
from culture to another, including on-line games and so forth
\cite{szell2012}, or during longer spans of an individual's life. We
are here looking at both of these quantitative characteristics with
the somewhat contradictory assumption that the persons described by
the data are "in an equilibrium". Thus, another important question
is: how does the eventual lack of stationarity matter here, and how
should it look like in the data? Life is not periodic, as it might
be in weekly or monthly patterns over a finite time span, and
moreover a society or culture is in a state of permanent flux. What
do the inevitable trends and transients look like, for a person? Are
there collective phenomena \cite{nokiafinconcl} quantified by the
entropy, patterns, and predictability computed over long but finite
windows in time - as during the times of a crisis in a society? Such
questions will need to be answered by access to much larger
data-sets.

\section*{Methods}
The location data consists of the raw cell ID timestamps and the
coordinates of the corresponding base-stations from a
smartphone-based application
\cite{raentocontext,raentosmartphones,eagle2009eigenbehaviors,zeit2011}.
The mapping of cell ID is difficult for issues inherent to the
mobile phone technology \cite{googlelocs}, for which reason we used
two methods (Text S1, Supplementary Figures S1-3):
 offline clustering
\cite{laasonen2009mining} and a fingerprint based method. This kind
of comparisons are useful since such approaches can not be perfect.
The original data set has more than 500 users, from a commercially
conducted ``user panel'' in southern Finland during 2008-10, before
processing. The participants (``panel members'') gave written
consent (``user agreement'') also concerning the gathering of the
data and its use for various purposes in the anonymized format,
including scientific studies. The location data is further
coarse-grained into time-slots of uniform length resulting in the
triplets (X,T,A) from above (see also
\cite{locchains,transitionslocation}). Here, we use the simplest
possible approach where the location A for the slot is chosen by a
majority rule from the detailed location information during the slot
length (see in general the Supplementary Figures S4-S11).

The data contains gaps, typically since the phone is closed down and
so is then the data gathering application (Supplementary Figures
S12-S14). The time-slots missing are partly covered by a padding
technique (user X closes the phone at home from 01.00 to 05.00 being
a typical case). Such gaps seem in general to contain some
information on the user behavior. Finally, days with more than 30~\%
of missing data after this procedure were entirely excluded,
reducing the missing data from initially almost 50~\% to less than
1~\%.  The slot length has been varied from 30 minutes to 2 hours,
out of which we show in this work representative results for the 1
hour case (recall the daily 24 hour cycle). The original data-set of
several hundreds of users is at the end pruned upto 66 persons, for
each of which at least 30 full days of data suitable for analysis
were found. In addition to the location data, we also have access to
a call and message data set for the same persons - except for the
"padded" time slots of course. This allows in principle to study the
correlations of user activity with both the communications patterns
and individual communication events such as text messages and calls,
including the anynomized phone numbers.

The patterns are resolved by a cluster-analysis (Supplementary
Figures S15-S26). The analysis is done to establish the typical days
or a set of characteristic behaviors. Note that we assume directly
that the human behavior here can be classified in this way : that
there is not simply a continuous spectrum of eigenbehaviors in
contrast to a discrete set \cite{eagle2009eigenbehaviors}, and that
it is reasonable to consider this on the particular level of
time-discretization (1 hour slot length). The easiest division here
might be "work-days" and "weekend-days - that separate to subclasses
as "Saturday" and "Sunday" - but a priori this is not given.

We use two methods to compare the influence of the technique;
naturally the result depends e.g. on the similarity metric used for
a pair of day-vectors. The first is $k$-means clustering and
Euclidean metrics on a binary expansion
(see Ref.~\cite{eagle2009eigenbehaviors}) of the data-set. The second method
is community detection on weighted graphs, where the days constitute
the vertices and the weights depend on the weighted Hamming distance
$d_h$ between any two as
\begin{equation}
 w_{ij}=\exp(-\beta d_h(i,j)),
\end{equation}
where $\beta$ is a parameter. We apply an extremal optimization
method (EO) ~\cite{duch2005community}, so that we can find the
number of clusters $k$ as a self-organized outcome. We compute a
single clustering, while for the $k$-means we do 200 runs with
random initial conditions.

\bibliography{locPLoS}
\newpage
\section*{Supporting Information Legends}

Supporting Information S1 contains information and further details
on the data handling and processing, and results (6 MBytes).

\section*{Figure Legends}

\clearpage

\clearpage

\clearpage

\end{document}